# Orbital mapping of energy bands and the truncated spin polarization in three-dimensional Rashba semiconductors

Qihang Liu[1,*], Xiuwen Zhang[1], J. A. Waugh[2], D. S. Dessau[1,2] and Alex Zunger[1,*]

[1]University of Colorado, Boulder, Renewable and Sustainable Energy Institute Colorado 80309, USA

[2]Department of Physics, University of Colorado, Boulder, Colorado 80309, USA

[*]Email: qihang.liu85@gmail.com, alex.zunger@colorado.edu.

## Abstract

Associated with spin-orbit coupling (SOC) and inversion symmetry breaking, Rashba spin polarization opens a new avenue for spintronic applications that was previously limited to ordinary magnets. However, spin polarization effects in actual Rashba systems are far more complicated than what conventional single-orbital models would suggest. By studying via first-principles DFT and a multi-orbital $k \cdot p$ model a 3D bulk Rashba system (free of complications by surface effects) we find that the physical origin of the leading spin polarization effects is SOC-induced hybridization between spin *and multiple orbitals*, especially those with nonzero orbital angular momenta. In this framework we establish a general understanding of the orbital mapping, common to the surface of topological insulators and Rashba system. Consequently, the intrinsic mechanism of various spin polarization effects, which pertain to all Rashba systems even those with global inversion symmetry, is understood as a manifestation of the orbital textures. This finding suggests a route for designing high spin-polarization materials by considering the atomic-orbital content.

The coupling between the motion of electrons and spins leading to spin polarization without external magnetic field is the focus of the emerging field of spin-orbitronics [1, 2], a branch of spintronics [3] that encompasses many interesting areas such as the Dresselhaus [4] and the Rashba [5] effects, spin-orbital toque [6, 7], topological insulation [8], and Majorana fermions [9]. The idea of control of spin degree of freedom even without external magnetic field is based on the fact that in a non- centrosymmetric system, spin-orbit coupling (SOC) sets up an effective internal magnetic field that creates spin splitting $E_1(\boldsymbol{k},\uparrow)-E_2(\boldsymbol{k},\downarrow)$ between spin-up and spin-down components in bands 1 and 2 away from the time-reversal invariant wavevector $K^*$. Specifically, Rashba spin splitting provides a classical scenario of spin topology encoded already in a simplified *single-orbital* (e.g., one *s* band) Hamiltonian $H = \hbar^2\boldsymbol{k}^2/2m^* + \lambda(\nabla V \times \boldsymbol{k})\cdot\boldsymbol{\sigma}$, where fully spin-polarized bands form two oppositely rotating spin loops at Fermi surface. However, in real materials spin can couple to *multiple* orbitals that make up the band eigenstates. The leading SOC spin effects that deviate from the classical single-orbital picture include: (i) the spin polarization $S_n(\boldsymbol{k})$ of each spin-split band (*n*, $\boldsymbol{k}$) appears to be truncated below its maximum value 100%; (ii) Each branch of the pair of spin-split bands (that are degenerate without SOC) experiences different degrees of spin truncation away from the time-reversal invariant wavevector $K^*$, resulting in a net *spin polarization* for the band pair (iii) The Rashba bands with two loops of energy contours can have identical helicities of spin texture. These effects were studied primarily in two-dimensional (2D) metallic surfaces — the classic Rashba systems [10-16]. Effect (i) and (ii) were theoretically discussed for freestanding Au(111) films [10], and in BiAg$_2$ metallic surface alloys [11, 16], while effect (iii) was predicted in the unoccupied bands of Bi/Cu(111) [12] and Bi/Ag(111) [14] surface alloys. The physics behind these intriguing spin effects was briefly touched upon in the context of 2D Rashba metallic films as being a consequence of the coupling between spin and different in-plane orbitals [12, 16]. However, in all 2D Rashba platform noted above there is a need for a free surface to observe the effects, so ordinary surface effects (such as broken bonds and surface band bending [17]) can cloud the intrinsic mechanism of these spin effects to be well established [13].

By extending the spin effects (i)-(iii) from the originally studied 2D metallic Rashba systems to 3D surface-free bulk Rashba semiconductors, we provide two pertinent generalizations: (a) We use the construct of *orbital texture* [a *k*-space map $I_n^{(l,m)}(k_x, k_y)$ of the content of orbital $m_l$ in band *n*, see Supplementary Eq. (S1)] familiar from topological insulators [18-22]. We point out that the effect of switch in orbital texture between two bands is common to topological insulators and to 3D Rashba semiconductors. Specifically, in the topologically-trivial bulk Rashba semiconductor the orbital textures of different Rashba bands switches from "radial" to "tangential" (with respect to the energy contour) character at the band crossing wavevector $K^*$, in full analogy with the phenomena previously observed [19, 22, 23] and calculated [21, 22] at the *surfaces* of 3D topological insulator (TI) $Bi_2Se_3$. Thus, the switch of orbital texture and indeed effects (i)-(iii) are not specific to topological or Rashba effects, but originate fundamentally from the fact that energy bands in complex solids invariably show a mixture of different azimuthal total orbital angular momentum (OAM) $m_j$, and that SOC can induce hybridization specifically between spin and *multiple* orbitals especially those with nonzero $m_l$, respectively. (b) We show that the spin polarization truncated by multiple orbital hybridization can be generalized even to systems with global inversion symmetry, manifesting the "hidden spin polarization effect" [24]. The understanding of effects (i)-(iii) and their reflection in the switch in orbital texture could provide better design guidelines for material selection and for spin manipulation in actual material application, e.g., electron confinement induced by spin-flip backscattering [16] and spin-galvanic effect [25, 26].

We reached these conclusions by applying density functional theory (DFT) and a multi-band *k* • *p* model to a 3D bulk Rashba compound BiTeI [27]. We find that within the six energy bands near the Fermi level ($E_F$) there are (i) large spin truncation per band at the band crossing wavevector $K^*$ with the residual spin polarization ranging from 0% to 85% far greater than the ~5 % seen in Au [111] surface; (ii) a net *spin polarization of band pairs* up to 50% for the top two valence bands, and (iii) identical spin-rotating loops at the occupied bands that can be examined by future angle-resolved photon emission spectroscopy (ARPES) measurements.

***Truncated spin polarization, net spin polarization and spin texture in BiTeI:*** This compound is a 3D bulk semiconductor [space group *P*3*m*1, see Fig. 1(a)] that manifest strong SOC, the ensuing orbital hybridization, and a polar field due to the positively charged Bi-Te layer that connects to the negatively charged I layer. The consequent Rashba spin-split bands [27, 28] from DFT calculations are shown in Fig. 1(b) (see Supplementary Materials for the DFT methods). We focus on the top four hole bands VB1-VB4 (going down from $E_F$) and bottom two electron bands CB1-CB2 (going up from $E_F$) around the $K^* = A(0,0,0.5)$ wavevector. The spin polarization $S_n(\mathbf{k})$ ($n$ = VB1-VB4, CB1-CB2) along $k_y$ direction is shown in Fig. 1(c). In what follows we discuss spin effects (i)-(iii) in bulk BiTeI:

(i) *Truncation of single band spin polarization:* The band-by-band spin polarization is calculated as the expectation value of the spin operator in each of the six spin-split bands at the wavevector $K^*$. We find that the magnitude of spin polarization is below the maximal magnitude of 100%. For VB1, VB2 and CB1, CB2 the spin polarization is ±85% and ±51%, respectively, while for VB3, VB4 the spin polarization is 0, i.e., a complete quenching of spin. Away from the band crossing wavevector $K^*$ the spin polarization of VB1 and VB2 evolves quite differently with $k_y$. VB1 is highly spin polarized in the considered momentum range up to $S_{VB1}$ = -96%. while VB2 loses its spin polarization rapidly with increasing $k_y$ down to $S_{VB2}$ = 46%.

(ii) *Net spin polarization of pairs of spin-split bands:* If we sum $S_{VB1}(\mathbf{k})$ and $S_{VB2}(\mathbf{k})$ (which would add up to zero in the absence of SOC), we find a *net* spin up of -50% at $k_y$ = 0.06 Å$^{-1}$. On the other hand, the sum of $S_{VB3}(\mathbf{k})$ and $S_{VB4}(\mathbf{k})$ reaches 48% at $k$ = 0.06 Å$^{-1}$, while $S_{CB1}(\mathbf{k})$ and $S_{CB2}(\mathbf{k})$ change slightly with $k$. We note that the position of the net polarization peak does not overlap with the energy peak in the band structure which locates at $k$ = 0.09 Å$^{-1}$ in $k_y$ direction, suggesting that the spin polarization is not a reflection of the eigenvalue dispersion.

(iii) *Identical directions of spin-rotation in the helical spin texture bands:* The band pair VB1+VB2 and the pair CB1+CB2 show the classical Rashba-type spin texture, i.e., opposite helicities of spin loops. However, as shown in Fig. 1(c) before $S_{VB4}$ falls below 0, $S_{VB3}$ and $S_{VB4}$ have the same sign, implying two spin loops with the same helicity in the area of $|k_{//}|$ < 0.10 Å$^{-1}$(see Supplementary Fig. S1-S3 the spin textures of all the six bands).

All three spin effects discussed above are absent in the conventional single-orbital model and thus reflect a manner of the complex interplay between spin and various OAM under the regime of SOC. To get a full picture it is useful to consider the *orbital texture*, i.e., the $k$-space map $I_n^{(l,m)}(k_x, k_y)$ of the content of orbital $m_l$ in band $n$.

***Orbital texture and its behavior for different bands:*** In real solids the orbital content generally varies with the wavevector and band index, reflecting the changing symmetry. Our base DFT calculation in this paper involves a vibrational calculation including *all* occupied states below $E_F$ in the solid via DFT total energy minimization, assuring physically realistic electronic structure. The orbital intensity is obtained by projecting the SOC-relevant band eigenstate ($n$, $\bm{k}$) onto local orbitals on atomic sites as shown in Supplementary Eq. (S1). This approach is different from the usual model Hamiltonian approaches that assume at the outset which orbitals will participate in given energy bands in a given crystal structure, notwithstanding the question if such orbitals would result in a total energy that supports the stability of the said crystal structure.

Figures 2(a)-2(d) show the DFT calculated orbital texture given by $p_y$ orbital intensity at different energy contours relative to $K^*$, for VB1 and VB2. We find that for VB1 the calculated $p_y$ orbital texture component is maximal along the $k_y$ direction and minimal along $k_x$ (where the $p_x$ orbital dominates the in-plane states). On the other hand, for VB2 the $p_y$ orbital texture component is minimal along $k_y$ and maximal along $k_x$. Thus, the orbital texture of VB1 and VB2 are different from each other and dominated, by *radial* and *tangential* in-plane orbital patterns, respectively. This difference leads to a radial-tangential orbital texture *switch*. To trace the switch between these two bands we follow Ref. [22] to define the in-plane orbital polarization $\lambda$ as a function of momentum $k$ and band index $n$ as $\lambda(n, \bm{k}) = \frac{I(p_x)-I(p_y)}{I(p_x)+I(p_y)}$, where $I(p_{x,y})$ denotes the calculated orbital intensity of $p_{x,y}$. Figure 2e shows $\lambda$ as a function of the in-plane azimuth angle $\theta$ (defined in Fig. 2b), confirming the switch of the intensity distribution in going from VB1 to VB2. Moreover, the intensity variation fits very well to a *sin2θ* or *cos2θ* distribution, with a period of π. As shown in Fig. 2f, $\lambda$ changes the sign as the momentum $k_y$, passing through $K^*$, indicating that the radial-tangential switch happens exactly at the band crossing wavevector. For VB3 and VB4 the orbital textures also have a switch between tangential

and radial characters at $K^*$ (see Supplementary Fig. S4). On the other hand, in CB1 and CB2 both Bi and Te atoms have considerable $p_{x,y}$ components, but with different orbital textures. For $p_{x,y}$ orbitals of Bi atom there is a radial-tangential switch from CB1 to CB2, while for $p_{x,y}$ orbitals of Te atom the orbital switch has an opposite trend, i.e., tangential-radial (see Supplementary Fig. S5). This observation agrees closely with the recent ARPES measurement by King et al. on the conduction surface state of BiTeI [29], and further confirms that such intriguing behavior comes from the intrinsic bulk state rather than any surface effects.

***Universality of orbital texture in bulk Rashba and surface of TI revealed by k • p modeling of DFT:*** The orbital texture switch between two bands at $K^*$ in the topologically-trivial semiconductor bares an interesting analogy to the recently observed angle-resolved photon emission spectroscopy (ARPES) measurements [19, 22] and DFT calculations [21] at the *surfaces* of TI $Bi_2Se_3$. Here we use a *multi-orbital k • p* model to illustrate how mixing orbitals of different $m_l$ and $m_j$ couple with spin and lead to the orbital texture switch and the spin polarization effects (i)-(iii). The crucial basis set represented in terms of $m_j$ is obtained in DFT; we now explicitly isolate it from all other DFT bands in $k \cdot p$ model below. Taking VB1 and VB2 as an example, we consider the SOC Hamiltonian as a perturbative form $H_R = \alpha(\sigma_y k_x - \sigma_x k_y)$ that is valid for both Rashba bulk and TI surface [21], and thus write the wavefunctions in the vicinity of $K^*$ as:

$$|VB1,k\rangle = \tfrac{1}{\sqrt{2}}(\sqrt{1-\omega_{VB}^2}+\mu_{VB}k)|p_t\rangle \otimes |LH\rangle + [\tfrac{i}{\sqrt{2}}(-\sqrt{1-\omega_{VB}^2}+\upsilon_{VB}k)|p_r\rangle + (\omega_{VB}-\xi_{VB}k)|Z\rangle] \otimes |RH\rangle \qquad (1)$$

$$|VB2,k\rangle = \tfrac{1}{\sqrt{2}}(-\sqrt{1-\omega_{VB}^2}+\mu_{VB}k)|p_t\rangle \otimes |RH\rangle + [\tfrac{i}{\sqrt{2}}(\sqrt{1-\omega_{VB}^2}+\upsilon_{VB}k)|p_r\rangle + (\omega_{VB}+\xi_{VB}k)|Z\rangle] \otimes |LH\rangle \qquad (2)$$

Where the in-plane $p$ orbital basis are tangential $|p_t\rangle = -\sin\theta |p_x\rangle + \cos\theta |p_y\rangle$ and radial $|p_r\rangle = \cos\theta |p_x\rangle + \sin\theta |p_y\rangle$; $|Z\rangle = \omega_s|s\rangle + \omega_z|p_z\rangle$ with $|\omega_s|^2 + |\omega_z|^2 = 1$; the spin basis are eigenstates of $H_R$, i.e., LH and RH helical spin states $|LH\rangle = \tfrac{1}{\sqrt{2}}\begin{pmatrix} ie^{-i\theta} \\ 1 \end{pmatrix}$ and $|RH\rangle = \tfrac{1}{\sqrt{2}}\begin{pmatrix} -ie^{-i\theta} \\ 1 \end{pmatrix}$; and $\omega_{VB}, \upsilon_{VB}, \xi_{VB}$ are the wavefunction coefficients that are band-dependent. More details on these wavefunctions can be found in Supplementary Materials. By calculating the difference of $p_t$ and $p_r$ intensity and omitting the higher-order k term we find that

$$[Int(p_t) - Int(p_r)]|_{VB1(VB2)} = \pm |k|[\sqrt{1-\omega_{VB}^2}(\mu_{VB}^* - v_{VB}^*) + c.c.] \quad (3)$$

From the modeling wavefunctions Eq. (1) and (2) and the difference of $p_t$ and $p_r$ intensity shown in Eq. (3), clear evidence is provided that the dominant in-plane orbital is different for the two spin-split valence bands, in other words the radial-tangential *orbital texture switches* at $k = 0$, i.e., the band crossing point $K^*$. The model also reveals that the symmetry lowering away from $K^*$ permits the mixing of new $m_j = \pm 3/2$ ($|+\rangle \otimes |\uparrow\rangle$ and $|-\rangle \otimes |\downarrow\rangle$, or "heavy hole" like) components into the $K^*$ wavefunctions ($m_j = \pm 1/2$ for the VB1 and VB2 of BiTeI), leading to the orbital texture switch.

Benefiting from the truncated basis set represented in terms of $m_j$, as distilled from the all-orbital DFT representation, we conclude that the underlying physical origin of the common behavior in both TI and non-TI materials is the SOC symmetry-enforced hybridization of different azimuthal total OAM $m_j$ components into band eigenstates.. This hybridization was absent at the high symmetry $K^*$ point. Thus, this effect is not limited to topological insulators but is far more general and applies also to Rashba compounds that are topological-trivial bulk semiconductors such as BiTeI [29] (see Supplementary Materials for more details on the comparison between Rashba bulk and TI surface).

***Understanding the spin polarization effects:*** Eq. (1) and (2) show that each orbital component couples with a certain spin state, forming *orbital-dependent spin textures* [see Supplementary Eq. (S2)]. Maximal spin magnitude arises when every orbital-dependent spin texture co-aligns, i.e., has the same helicity. This requires that the band eigenstate be composed exclusively of orbitals with the same azimuthal quantum number $m_l$. In real materials where SOC mixes orbitals with different $m_l$ in one eigenstate ($n$, **k**), the corresponding spin polarization is truncated relative to its maximal value. Specifically, the tangential in-plane orbital ($p_t$) always couples opposite spin texture to that of radial in-plane orbital ($p_r$), s and $p_z$ orbital. At the wavevector A with $k \rightarrow 0$, $p_t$ and $p_r$ components have the same intensity but opposite spin pattern and thus cancel each other, making the spin polarization $S(k\rightarrow 0) = |\omega_{VB}|^2$ all contributed by $s$ and $p_z$ orbital. This scenario gives the truncated spin polarization at $K^*$ for all the bands shown in Fig. 1f. Especially, VB3 and VB4 are $m_j = \pm 3/2$ states with the corresponding wavefunctions at $K^*$ containing only in-plane $p$ components: $|p_+\rangle\otimes|\uparrow\rangle$ and $|p_-\rangle\otimes|\downarrow\rangle$, leading to the complete quenching of spin. The total spin polarization summing over all bands is equivalent to the value obtained from the contributions of $m_l = 0$ states, e.g., $s$, $p_z$, and $d_z^2$, etc. This statement is valid also in the traditional 2D Rashba systems like Au(111) surface [10], in which the surface Rashba bands are nearly exclusively composed by the $s$ and $p_z$ states, and thus have nearly 100% spin polarization.

Due to the orbital texture switch, we find from Eq. (1) and (2) that the dominat in-plane $p$ orbitals of a pair of Rashba bands couple to spin textures with the same helicity. This fact is confirmed by DFT calculation showing that the dominating radial orbital for VB1 and tangential orbital for VB2 both have RH spin texture (see Fig. 3). Therefore, by summing the in-plane orbital contribution and the $s+p_z$ orbital contribution to the spin textures, the total spin polarization shows different degree of truncation for VB1 and VB2, as shown in Fig. 1f (see Supplementary Materials for more details on orbital-dependent spin textures for different bands and the effects of SOC strength). For VB3 and VB4, $p_t$ and $p_r$ dominate the whole state around $K^*$, respectively. Consequently, the total spin textures form two LH helical spin loops (see Fig. S3 and S4). Unlike the case in Bi/Cu(111) [12] and Bi/Ag(111) [14] surface alloys, these identical spin-rotating loops occur at the occupied bands that is detectable by ARPES measurements.

*Generalizing to systems with inversion symmetry:* The truncated spin polarization and the net polarization effect are illustrated above in a Rashba semiconductor BiTeI. However, we are not using any special feature of this orbitally-hybridized compound other than SOC-induced (Rashba) spin splitting, and thus expect our finding to pertain to a very broad range of such compounds, whether inversion symmetry is present or not. In centrosymmetric crystals the spin bands are degenerate in $E$ vs $k$ momentum space, but this does not mean that the spins are mixed in position space. Correspondingly, the centrosymmetric systems with lower-symmetry sectors could manifest a "hidden" form of spin polarization named R-2 effect [24, 30]. Like two oppositely stacked Rashba layers, in R-2 system there is still finite spin polarization localized on each atomic site *i* that feels inversion-asymmetric environment, and such polarization is compensated in *k* space by another atom forming inversion partner of site *i*. Therefore, if we consider the local spin polarization in real space, i.e., localized on one inversion-asymmetric atoms or sectors, the spin truncation effects also happen when spin is coupled by multi-orbitals with different $m_l$. We choose a centrosymmetric R-2 material LaOBiS$_2$ (using the reported space group *P4/nmm*) to illustrate the truncation effects and find that the hole spin has the spin polarization ~90%, while the electron spin is only ~30% polarized (see Supplementary Materials for more details). It is noticeable that the orbital-dependent spin texture is robust again small perturbation that breaks the global inversion symmetry such as electric fields, which is highly feasible for the detection on experiments.

***Discussion and design implications:*** In the past few years wide areas of physics and material science that related to SOC build up the new field spin-orbitronics. By generalizing the previously observed orbital texture switch in TI surface to a bulk Rashba semiconductor, we unveil the deeper mechanism of various spin polarization effects that is unexpected in the simple Rashba model, and provide a clear picture of the delicate interplay between spin and multi-orbitals. Our work is also expected to open a route for designing by the atomic-orbital feature high spin-polarization materials that are of vital importance for nonmagnetic spintronic applications [3]. For example, in Rashba splitting the in-plane spin drives a current perpendicular to the direction of spin polarization induced by the asymmetric Elliot-Yafet spin-relaxation, named spin-galvanic effect [25, 26]. Since the current is proportional to the average magnitude of the spin polarization, the effects of spin truncation and that different branches experience different degrees of spin polarization could have a more complex impact on the conversion process between spin and current in a Rashba system, which calls for further investigation.

**Acknowledgement**

We are grateful for helpful discussions with Dr. Yue Cao. This work was supported by NSF Grant No. DMREF-13-34170. This work used the Extreme Science and Engineering

Discovery Environment (XSEDE), which is supported by National Science Foundation grant number ACI-1053575.

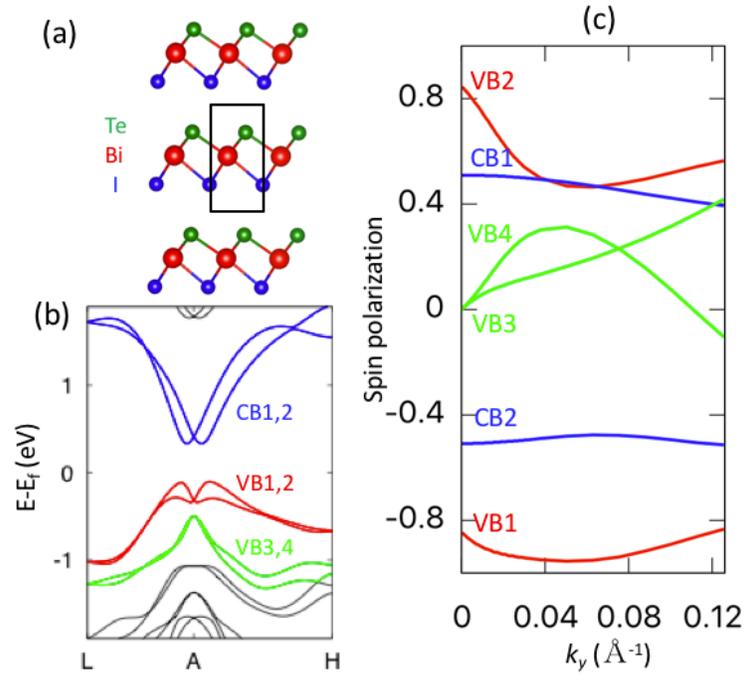

**Fig. 1:** (a) Crystal structure and (b) DFT band structure of BiTeI along the high-symmetric line L(0,0.5,0.5) – A(0,0,0.5) – H(1/3,1/3,0.5). (c) Spin polarization $S_n(\mathbf{k})$ of the six bands VB1-4 and CB1,2 along $k_y$ direction at the $k_x = 0$ cut. Note that Spin polarization at $k_y < 0$ fulfills $S_1(-\mathbf{k}) = -S_2(\mathbf{k})$ due to Kramer's degeneracy.

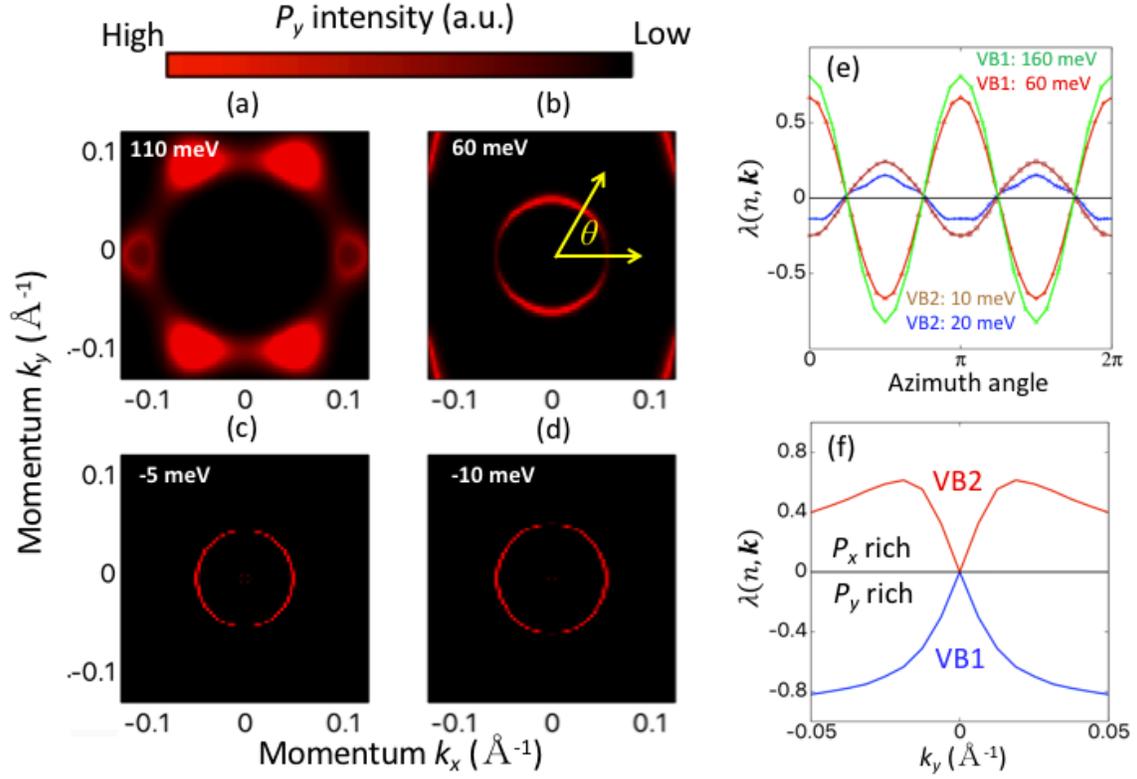

**Fig. 2:** (a-d) Orbital texture indicated by $p_y$ intensity at different energy contours relative to the band crossing point, (a,b) for VB1 and (c,d) for VB2. (e,f) In-plane orbital polarization $\lambda$ for (a) different energy contours as a function of the azimuth angle $\theta$ defined in panel b, and for (f) different spin-splitting bands as a function of the momentum $k_y$ at the $k_x = 0$ cut. Note that the orbital polarization switch signs exactly at the band crossing point $k_y = 0$.

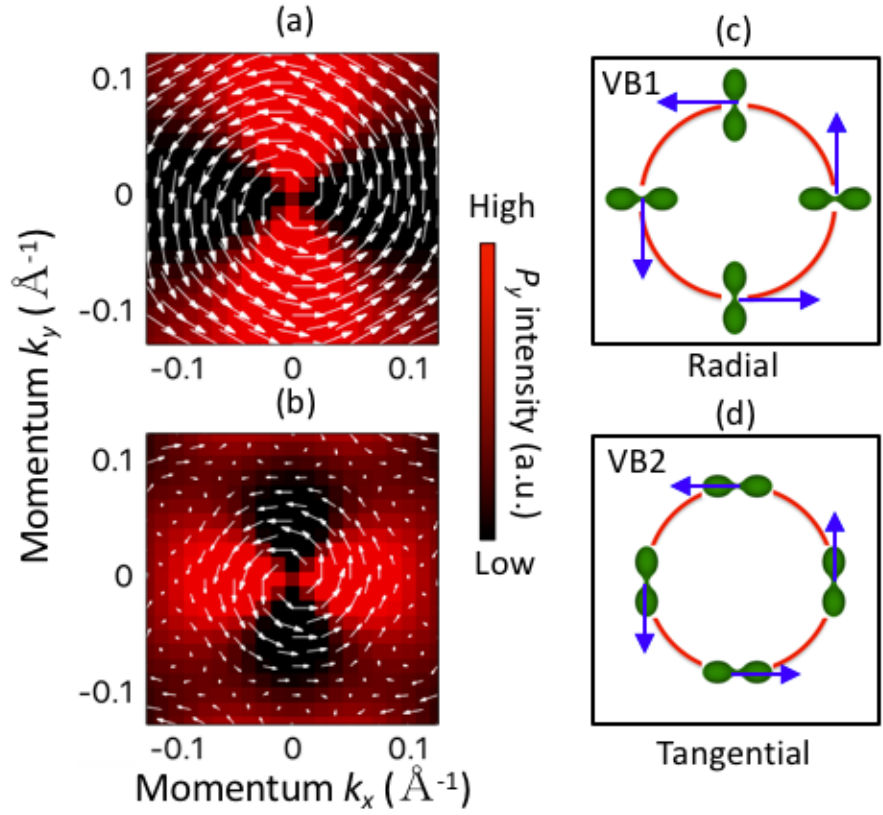

**Fig. 3:** (a,b) The in-plane *p*-dependent spin textures (white arrows) for VB1 (a) and VB2 (b). The background color indicates the $p_y$ intensity. (c,d) Schematic plots for the radial orbital texture and the corresponding orbital-dependent spin texture of VB1 (c) and the tangential orbital texture and the corresponding orbital-dependent spin texture of VB2 (d).